\documentstyle[12pt]{article}
\def\be{\begin{equation}}
\def\ee{\end{equation}}
\def\bea{\begin{eqnarray}}
\def\eea{\end{eqnarray}}

\begin{document}

\begin{center}
{\Large{\bf Generalization of Some Algebras in the Bosonic
String Theory}}

\vskip .5cm {\large Seyed Sina ShahidZadeh Mousavi} \vskip .1cm
 {\ Faculty of Physics, Amirkabir University of Technology (Tehran
 Polytechnic)}
\\  P.O.Box: 15875-4413, Tehran, Iran \\
{\sl e-mail: shahidzadeh@aut.ac.ir}
\\\end{center}

\begin{abstract}

We assume that the total target phase space is non-commutative. This
leads to the generalization of the oscillator-algebra of the
string, and the corresponding Virasoso algebra. The effects of this
non-commutativity on some string states will be studied.

\end{abstract}

{\it PACS numbers}: 11.25.-w

{\it Keywords}: Bosonic string; Non-commutativity.

\vskip .5cm
\newpage

\section{Introduction}

The non-commutative geometry \cite{1} has been considered for some time in connection with
various physics subjects. Recent motivation to study the non-commutative geometry mainly
comes from the string theory. String theories have been pointing towards a non-commuting scenario already in
the 80's \cite{2}. Various subjects in the non-commutativity in string theory
can be found in the Refs.\cite{3,4,5,6,7,8,9}.
We are interesting to the non-commutative world-sheet of the bosonic string, e.g. see Ref.\cite{9}. Thus, we
 study the string propagation in the non-commutative phase
space. That is, we consider the following commutation relations
 \bea
&~&[X^\mu(\sigma,\tau),\Pi^\nu (\sigma^\prime,\tau)
]=i\eta^{\mu\nu} \delta(\sigma-\sigma^\prime ),
\nonumber\\
&~&[X^\mu(\sigma,\tau),X^\nu (\sigma^\prime,\tau)
]=i\theta^{\mu\nu}(\sigma-\sigma^\prime ),
\nonumber\\
&~&[\Pi^\mu(\sigma,\tau),\Pi^\nu (\sigma^\prime,\tau)
]=i\gamma^{\mu\nu}(\sigma-\sigma^\prime ).
 \eea
where $\Pi^\mu(\sigma,\tau)=\frac{1}{2
\pi\alpha^\prime}\partial_\tau X^\mu(\sigma,\tau)$ is the canonical momentum, conjugate to $X^\mu(\sigma,\tau)$. The variable
$\theta^{\mu\nu}$ indicates the non-commutativity of the space
part, and $\gamma^{\mu\nu}$ shows the non-commutativity of the
momentum part of the phase space.

This paper is organized as follows. In section 2, the generalized oscillator algebra will be obtained. In section 3,
the associated Virasoro algebra will be studied. Section 4 is devoted to the conclusions.

\section{Oscillator algebra}

The Fourier expansions of the variables
$X^\mu(\sigma,\tau),$ $\Pi^\mu(\sigma,\tau),$ $\theta^{\mu\nu}(\sigma-\sigma^\prime
)$ and $\gamma^{\mu\nu}(\sigma-\sigma^\prime )$ are as in the following
\bea
\theta^{\mu\nu}(\sigma-\sigma^\prime
)=\sum_{n=-\infty}^{\infty}\theta_n^{\mu\nu}
e^{in(\sigma-\sigma^\prime)}, \eea
\bea
\gamma^{\mu\nu}(\sigma-\sigma^\prime
)=\sum_{n=-\infty}^{\infty}\gamma_n^{\mu\nu}
e^{in(\sigma-\sigma^\prime)}, \eea
\bea
 X^\mu (\sigma,\tau)=x^\mu+{2\alpha^\prime } p^\mu
\tau+i\sqrt{2\alpha^\prime }
\sum_{n\neq0}^{\infty}\frac{1}{n}\alpha_n^\mu \cos{n\sigma}
e^{-in\tau}, \eea where, for simplicity we consider the open
string solution.

Introducing the mode expansions (2)-(4) is the equations (1)
gives the following oscillator-algebra
\bea &~&[p^\mu,p^\nu
]=i\pi^2 \gamma_0^{\mu\nu},
\nonumber\\
&~&[x^\mu,p^\nu ]=i\eta^{\mu\nu}-2i\pi^2
\alpha^\prime\tau\gamma_0^{\mu\nu},
\nonumber\\
&~&[x^\mu,x^\nu ]=i\theta_0^{\mu\nu}-4i\pi^2 \alpha^{\prime2}
\tau^2\gamma_0^{\mu\nu}.
 \eea
 for the zero-modes, and
 \bea
 [\alpha_m^\mu,\alpha_n^\nu ]=\bigg{(}m\eta^{\mu\nu}+2i\alpha^\prime \pi^2 \gamma_n^{\mu\nu}+\frac{n^2}{2\alpha^\prime}\theta_n^{\mu\nu} \bigg{)}
 \delta_{n+m,0},
 \eea
 for the oscillating-modes.

 We observe that the non-commutativity of the
 phase space modifies the algebra. In other words, even if its space
 part is commutative , $ i.e.$  $\theta^{\mu\nu}=0$, the parameter $\gamma_0^{\mu\nu}$ tells us
 that zero-modes of the space part is non-commutative. However, the
 oscillating algebra is affected by both non-commutativity parameters $\theta^{\mu\nu}$ and
 $\gamma^{\mu\nu}$.
 Vanishing $\theta^{\mu\nu}$ and
 $\gamma^{\mu\nu}$ implies the usual algebra for the string
 modes, as expected.

\subsection{Conditions on the non-commutativity parameters}

Now take the Hermitean conjugate of the both sides of the second
and third equations of (1), This leads to the equations \bea
[\theta^{\mu\nu}(\sigma-\sigma^\prime)]^\dagger=-\theta^{\nu\mu}(\sigma^\prime-\sigma),
\eea \bea
[\gamma^{\mu\nu}(\sigma-\sigma^\prime)]^\dagger=-\gamma^{\nu\mu}(\sigma^\prime-\sigma),
 \eea
 In terms of the oscillating modes, we obtain
  \bea
&~&(\gamma_n^{\mu\nu} )^\dag=-\gamma_n^{\nu\mu},
\nonumber\\
&~&(\theta_n^{\mu\nu} )^\dag=-\theta_n^{\nu\mu}. \eea That is,
effect of the Hermitean conjugation from changing the mode index $"n"$
of $\alpha_n^{\mu}$ has been modified to the exchange of the
space-time indices $\mu$ and $\nu$ .

\section{The corresponding Virasoro algebra}

We only assumed the quantization (1). Therefore, the string action
does not change. This implies that the Virasoro operators remain as previous, $i.e.$,
\bea
L_m^{(\alpha)}=\frac{1}{2}\sum_{n=-\infty}^{\infty}\alpha_{m-n}^\mu
\alpha_{n\mu}. \eea Due to the modification of the
oscillator-algebra (6), the corresponding Virasoro algebra also
is modified, $i.e.$, \bea
 [L_m^{(\alpha)},L_n^{(\alpha)} ]=(m-n) L_{m+n}^{(\alpha)}+\frac{d}{12}
m(m^2-1) \delta_{m+n,0}+\pounds_{mn}, \eea where $\pounds_{mn}$ is
the consequence of the non-commutativity \bea
\pounds_{mn}=\frac{1}{2}
\sum_{\ell=-\infty}^{\infty}\lambda_{mn,\mu\nu}
\alpha_\ell^\nu \alpha_{m+n-\ell}^\mu,
\eea
 \bea \lambda_{mn}^{\mu\nu}=2i\alpha^\prime
\pi^2 (\gamma_{n-m}^{\nu\mu}+\gamma_{n-m}^{\mu\nu}
)+\frac{(n-m)^2}{2\alpha^\prime}\bigg{(} \theta_{n-m}^{\nu\mu}+
\theta_{n-m}^{\mu\nu} \bigg{)}. \eea
Therefore, the second and third terms of the right-hand-side of (11) are originated from the anomaly.

  \subsection{The ghosts contribution}

  Introducing the conformal ghosts $b(\sigma,\tau)$ and
  $c(\sigma,\tau)$, the Virasoro operator takes the form
  \bea
  L_m=L_m^{(\alpha)}+L_m^{(g)},
  \eea
  where
  \bea
  L_m^{(g)}=-\sum_{n=-\infty}^{\infty}(m-n) b_{m+n} c_{-n}.
  \eea
  According to the anti-commutation relation ${\{c_m,b_n\}}=\delta_{m+n,0}$ the
  Virasoro algebra of $L_m$ becomes
  \bea
  [L_m,L_n ]=(m-n) L_{m+n}+{{\frac{d-26}{12}m( m^2-1)
  }}\delta_{m+n,0}+\pounds_{mn}
  \eea
 Therefore, for the choice $d=26$ the usual
 anomaly is removed. However, the $\pounds_{mn}$ which is anomaly due to the non-commutativity always remains.

  \section{Conclusions}

Without modification of the string action, we assumed a total non-commutativity target phase space. This non-commutativity
 is induced to the oscillator algebra. Thus
, we have a modified algebra. The non-commutativity of the momentum part implies that the total zero-modes of the phase space
become non-commutative. However, these non-commutativity parameters are restricted by some conditions.

The Virasoro operators save their forms, as in the commutative case. The modification of oscillator-algebra induces
an extra anomaly term in the Virasoro algebra.



\begin{thebibliography}{99}
 \bibitem{1}
 A. Connes, {\it "Noncommutative Geometry, Academic Press"}
  (1994) .
 \bibitem{2}
 E. Witten, Nucl. Phys. B268(1986)253.
 \bibitem{3}
 N. Seiberg and E. Witten, JHEP 9908(1999)032, hep-th/9908142.
   \bibitem{4}
  A. Connes, M.R. Douglas and A. Schwarz, JHEP 9802(1998)003, hep-th/9711162.
 \bibitem{5}
 Y.E. Cheung and M. Krogh, Nucl. Phys. B528(1998)185, hep-th/9803031; A. Schwarz,
 Nucl. Phys. B534(1998)720, hep-th/9805034, C. Hofman and E. Verlinde, JHEP
9812(1998)010, hep-th/9810116.
\bibitem{6}
  M.R. Douglas and C. Hull, JHEP 9802(1998)008, hep-th/9711165.
 \bibitem{7}
 E. Fradkin and A. Tseytlin, Phys. Lett. B158(1985)316; C.S. Chu and P.M. Ho, Nucl.
 Phys. B550(1999)151, hep-th/9812219; Nucl. Phys. B568(2000)447, hep-th/9906192;
 V. Schomerus, JHEP9906(1999)030, hep-th/9903205
 \bibitem{8}
 A. Fayyazuddin and M. Zabzine, Phys. Rev. D62(2000)046004, hep-th/9911018; P.M.
 Ho and Y.T. Yeh, Phys. Rev. Lett. 85(2000)5523, hep-th/0005159; D. Bigatti and L.
 Susskind, Phys. Rev. D62(2000)066004, hep-th/9908056; N. Ishibashi, hepth/
 9909176.
   \bibitem{9}
  D. Kamani, Eur. Phys. J.C. 26(2002)285, hep-th/0008020; Mod. Phys. Lett. A 19(2004)375, hep-th/0102141.


\end{thebibliography}
\end{document}